\documentclass[11pt]{article}
\usepackage[utf8]{inputenc}
\usepackage{indentfirst}
\usepackage{amsmath}
\usepackage{amssymb}
\usepackage{url}
\usepackage[colorlinks=true,linkcolor=blue,urlcolor=blue,citecolor=blue,anchorcolor=blue]{hyperref}
\usepackage{graphicx}
\usepackage[font=footnotesize,labelfont=bf]{caption}
\usepackage[letterpaper, margin=1in]{geometry}

\usepackage{cite}

\title{Group adaptation drives opinion dynamics in higher-order networks
}

\author{Cosimo Agostinelli, Marco Mancastroppa, Alain Barrat\\
\small \textit{Aix-Marseille Univ, Université de Toulon, CNRS, CPT,}\\ 
\small \textit{Turing Center for Living Systems, 13009 Marseille, France}
}

\begin{document}
\maketitle

\begin{abstract}
In modern interconnected societies, opinions and beliefs can quickly spread across large populations, giving rise to collective behaviors such as the adoption of social norms or polarization. The interest in these phenomena led to the formulation of many models 
aiming at reproducing relevant emergent properties from simple mechanisms of interactions
between individuals. In particular, opinion dynamics models
mimic how the opinions of individuals on a given topic may evolve when they interact,
and study the conditions for global consensus or polarization. Most models assume that these
interactions occur between pairs of agents, typically on a fixed network structure.
However, discussions leading to opinion changes can occur in groups, and these groups can also undergo adaptive changes and modifications if their members disagree. 
Here, we propose a bounded confidence model of opinion dynamics taking into account these
two mechanisms: a group discussion can lead to a global agreement among all group members,
while a strong divergence of opinions within a group leads to its splitting, 
followed by merging of the resulting subgroups with other groups.
We systematically study the outcome of this model as a function of the tolerance of agents for reaching an agreement. 
Strikingly,
adaptivity seems to suppress important effects induced by group interactions, and 
to restore a phenomenology close to the one obtained with pairwise interactions.
We show that adaptivity, which allows the formation of large groups, prevents the transition to a fragmented state at small tolerance. Moreover, it restores a phase transition from a polarized state to consensus, which would otherwise disappear due to group effects in a non-adaptive bounded confidence model with group interactions.
Overall, our work shows that both adaptivity and group interactions shape the structure
of social ties and the global opinion dynamics in a population. 
\end{abstract}

\section{Introduction}
Understanding how opinions evolve within a population is an important task in modern social sciences \cite{Haoxiang2011,Noorazar2020b,Castellano2009,starnini2025,caldarelli2025physics}.
In this context, a broad variety of opinion dynamics models have been developed 
to mimic mechanisms through which individuals influence each other, and how collective behaviors, ranging from polarization to consensus formation, can emerge from simple rules of interactions 
inspired from social mechanisms
\cite{Castellano2009,starnini2025,caldarelli2025physics,deffuant2000, shirzadi2025, peralta2022}.

Among these, bounded confidence models form a well-established and widely studied class of models 
\cite{deffuant2000,Lorenz2007, Noorazar2020a, Liang2013, Brooks2024}:
they describe individuals’ opinions as real-valued variables and assume that
opinion changes can only occur through interactions, when the difference between the opinions of interacting
individuals are smaller than a prescribed confidence threshold (the parameter of the model).
In this case, an interaction between two individuals lead them to bring their opinions
closer.
Despite the simplicity of their definition, bounded confidence models reproduce a rich phenomenology, including clustering, polarization, and the emergence of global consensus: 
the microscopic rules governing interactions translate into collective patterns that are  sensitive to the confidence bound, the distribution of initial opinions, and also to the structure of interpersonal connections \cite{Lorenz2007,Brooks2024,porter2018}. Indeed, the original model formulation \cite{deffuant2000}
considers interactions between randomly chosen pairs of individuals, while more realistic
setups consider that agents' interactions can be represented as taking place 
along a social network, where nodes correspond to individuals and links capture the possibility of a direct influence or communication event \cite{barrat2008dynamical,Castellano2009,Liang2013}.
This framework allows the study of how the network topology shapes the macroscopic outcomes of opinion formation \cite{porter2018,MF_Deffuant2021,MF_Deffuant2023}. 
However, real-world interactions are not limited to pairwise relations: many social activities--such as group discussions, committee decisions, or collaborative online exchanges--naturally involve interactions among more than two individuals at a time.
Empirical and theoretical evidence suggests that explicitly taking into account such group 
(higher-order) interactions, by using models defined on hypergraphs, can better represent
the complexity of real-world social influence processes \cite{Levendusky2016, battiston2020, battiston2025} and result in dynamical behaviors that have no correspondence in purely pairwise settings \cite{Iacopini2019,Arruda2020, Iacopini2022, civilini2024}. 
In particular, several studies in the context of opinion dynamics have shown that higher-order interactions can substantially alter the system’s phenomenology, leading to qualitative shifts in the onset of consensus, the stability of opinion clusters, and the pathways through which collective agreement is reached \cite{Hickok2022, Neuhauser2020, Sahasrabuddhe2021,Schawe2022}.

These works consider fixed interaction structures between individuals.
However it is also  
important to take into account that the evolution of individual opinions can itself
result in a modification and change in these structures, through system adaptation to the opinion dynamics. To model such feedback effects, adaptive
network structures can be used, in which connections co-evolve with opinions \cite{Holme2006, kozma2008, Kan2023}: for instance, individuals may break ties with neighbors holding opinions outside their confidence range, leading to a dynamic reshaping of the interactions.
When interactions are pairwise, it has been shown that this interplay can 
significantly alter the model's properties \cite{Holme2006, kozma2008,Nardini2008, Kan2023}.
Adaptive behavior can however also concern group interactions: 
individuals may withdraw from group discussions, or groups can split into smaller
subgroups if disagreement is too strong, and isolated
individuals or small groups may also join other groups \cite{Iacopini2024}, searching for similar opinions. 
The impact of such higher-order adaptivity, in which the topology of group structures becomes coupled to the evolving opinions, has been studied only for few cases \cite{Horstmeyer2020, Burgio2025,mancastroppa2026adaptive,mancastroppa2026adaptive2}.


Here, we contribute to this effort by proposing a bounded confidence model that incorporates both adaptivity and group interactions. Specifically, we generalize the Deffuant model of opinion dynamics with group interactions \cite{deffuant2000,Schawe2022} to take into account
adaptive behaviors: we assume that members of a group can converge on a common opinion, if  
their opinions are close enough, but in the opposite
case the group splits into smaller subgroups. These subgroups can in turn rearrange their
connections and merge with other groups.
We study how this interplay of the dynamics of both opinions and group structure affect the
model properties, as a function of the confidence parameter.
We show that the model's phenomenology is very close to the one of
the adaptive Deffuant model with only pairwise interactions, with no fragmented phase
at small tolerance, and a clear transition from a polarized state to a consensus state. 
This is in stark contrast with the non-adaptive model's properties, which are strongly 
modified when  higher-order interactions are taken into account \cite{Schawe2022}. 
Our results thus hint that higher-order effects in dynamical processes are limited
when structure adaptivity is taken into account, as it was also shown recently for spreading processes
in \cite{mancastroppa2026adaptive}. 


\section{The model}

We consider a population of N agents who interact by comparing their opinions about a given topic.
As in the original Deffuant model \cite{deffuant2000}, the opinions of agents $i=1,...,N$ are represented by real numbers $x_i \in [0,1]$ and interactions happen in discrete time steps $t=1,...,T$.
For example, $x_i$  may denote the political orientation of an individual
confronted to a choice of two opposing parties. 
In this case, extreme values (close to 0 or 1) correspond to strong alignment with one party, while intermediate values indicate milder opinions and uncertain agents.
At each time step of the model, agents interact and compare their opinions with their neighbours. 
If their opinions are close enough,
namely, if they differ less than a given threshold $\varepsilon$,
they reach an agreement and their opinions are updated to a common value.
Otherwise a structural adaptive modification is triggered.
The key parameter is the \textit{confidence} $\varepsilon$, which encodes how 
tolerant the agents are and how much they are willing to adjust their opinions.
In the original model \cite{deffuant2000}, agents interact in a pairwise manner. 
Here instead, we take into account that 
interactions in realistic settings can occur in groups of arbitrary size, 
by representing these interactions as a hypergraph $\mathcal{H}$, where nodes correspond to agents and hyperedges to group interactions \footnote{In the following we will use equivalently the terms nodes and agents, as well as group and hyperedge (or hyperlink).}\cite{battiston2020,Schawe2022}.
Comparison of opinions therefore takes place in each group: if the group members
hold sufficiently close opinions, the discussion leads the whole group to an agreement  (all agents' opinions in the group take the same value);
otherwise, the hyperedge representing that particular interaction splits into multiple subgroups. 
Afterwards, each of these subgroup may merge with other groups. 
These steps describe thus a coupled evolution of the structure of interactions, through group splitting and merging, 
and of opinions, through group agreement. 

\begin{figure}[t]
    \centering
    \includegraphics[width=\textwidth]{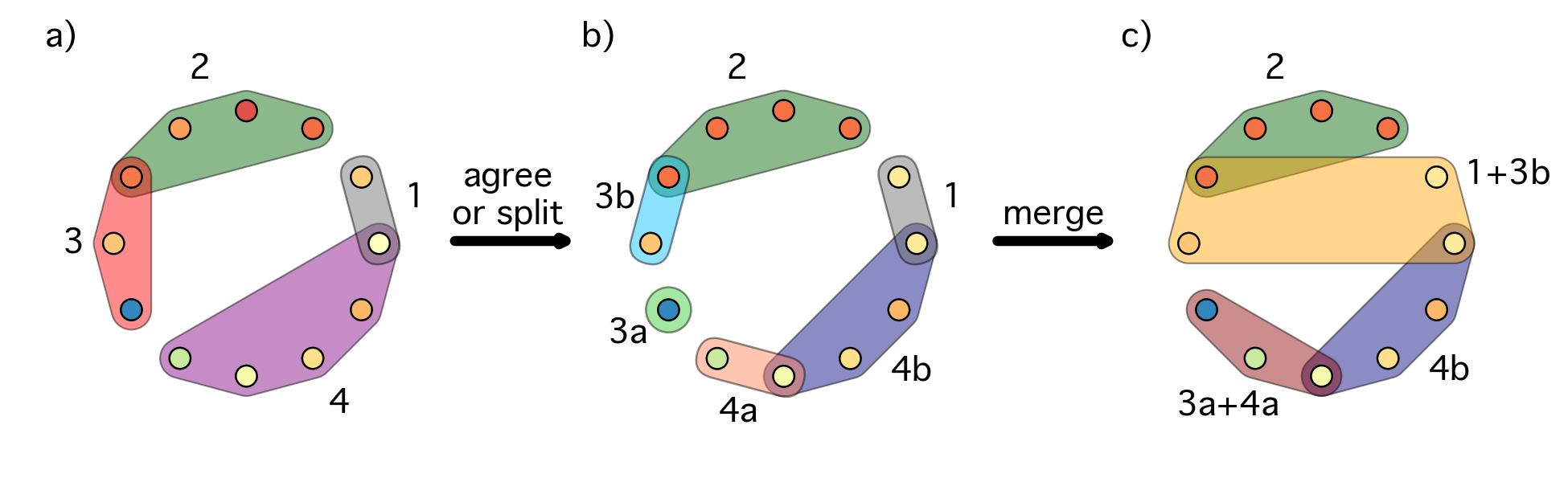}
    \caption{\textbf{Illustration of the two phases composing one time step of the model.}
    Nodes (agents) are colored according to their opinion and each hyperedge (group) is identified by a number and a different color.
    \textbf{(a)} Initial configuration composed by four groups. 
    \textbf{(b)} Agents forming group 1 and group 2 converge respectively to a common opinion. Group 4 instead splits into two overlapping subgroups (4a and 4b), while group 3 divides into two disjoint subgroups (3a and 3b, with 3a being an isolated node). 
    \textbf{(c)} Several of the subgroups resulting from the splits of groups 3 and 4 merge with other hyperedges, generating new groups (3a+4a and 1+3b).
    }
    \label{fig: sketch}
\end{figure}

More in detail, we start from a hypergraph $\mathcal{H}_0$ encoding the initial set of interactions among $N$ agents (nodes).
We assign to each individual an initial opinions $x_1,...,x_N$ and we perform 
at each time the following steps (see also Fig. \ref{fig: sketch} for a schematic representation).

\textit{1. Agreement or split.} In this first phase,
a hyperedge $e$ is randomly selected and the opinions of nodes in $e$ are compared.
If:
\begin{equation}
    \max_{i \in e} x_i - \min_{i \in e} x_i < \varepsilon , 
    \label{eq:max_min_condition}
\end{equation}
i.e., if the dispersion of opinions is not too large,
all agents in $e$ converge to the average opinion, i.e., 
$x_i \leftarrow \langle x_j \rangle_{j \in e} \ \ \forall i\in e$.
If the condition of Eq. \eqref{eq:max_min_condition} is not satisfied, convergence does not occur: 
the nodes retain their current opinions and $e$ is split into several subgroups.
Each subgroup $e'$ is built in order to contain only nodes whose opinions are mutually within distance $\varepsilon$, mimicking how real discussions lead to group fragmentation 
when disagreement is too strong.
We proceed as follows to build the subgroups of $e$:
\begin{itemize}
    \item[i.] we choose uniformly at random a node $i$ in $e$ to serve as seed of a subgroup;

    \item[ii.] we collect all nodes $j,k,... \in e$ whose opinions differ from that of $i$ by less than $\varepsilon$;
    
    \item[iii.] we form a new hyperedge $e'$ consisting of these nodes;
    
    \item[iv.] we repeat the procedure selecting the seeding node from $e \smallsetminus e'$ to create $e''$; then from $e \smallsetminus (e' \cup e'')$ to create $e'''$, etc until no further seeding nodes can be selected.
    
\end{itemize}
Note that at each step of the iteration, a node $j$ included already in a subgroup (such as in $e'$, $e''$, etc) cannot serve as seed of a new subgroup in step iv., but it
can still be included in another subgroup of seed $k$ if $x_j$ and $x_k$ differ less than $\varepsilon$. Such nodes can thus act as ``bridges” between different
subgroups resulting from the split of the original hyperedge $e$ (see Fig. \ref{fig: sketch}a-b), which can hence present some level of overlap.
We also note that this process can result in some isolated nodes.

We apply the procedure of agreement or split to all hyperedges of the hypergraph, taken in random succession, and we then proceed to the second phase of the time step.

\textit{2. Merger.} In this phase, each subgroup resulting from the splitting phase can
merge with other groups, and we assume that the probability to merge depends on the groups' 
sizes. The rationale is that small subgroups tend to seek interaction partners to avoid isolation, while larger groups are less likely to do so.
We model this  higher propensity  of small groups to merge by assigning to each subgroup $e'$ a probability $p(e') = 1/|e'|$ of attempting to merge with another existing hyperedge.
The target hyperedge $e_t$ that $e'$ joins can be either another subgroup
resulting from a split, or a group that reached consensus in the previous phase. 
It is thus selected among all hyperedges in the hypergraph resulting from the first phase of the time step (Fig. \ref{fig: sketch}b-c), with probability depending on the
size of the merged group, $q(e' \rightarrow e_t) \propto 1 / |e' \cup e_t|$. 
The motivation behind this condition is that mergers leading to very large groups are less likely, consistently with empirical observations on face-to-face group interactions \cite{Cencetti2021, Iacopini2024}.
Importantly, we do not require members of $e'$ to have any prior information about the opinions of agents in $e_t$ \cite{kozma2008,Kan2023}, hence the merging is only driven by the groups' sizes and not by the opinions within them.
We finally note that we do not take into account a possible multiplicity of hyperedges: if an already existing group $e$ is created anew by either a split or a merge, we ignore it.

We let the model evolve according to the rules mentioned above until 
the following convergence criterion is satisfied \cite{Schawe2022}, indicating a
steady state in which opinions do not evolve anymore:
\begin{equation}
    \sum_{i=1}^N |x_i(t+1) - x_i(t)| < 10^{-3} \ .
    \label{eq:conv_crit}
\end{equation}
This criterion by itself does not ensure that the 
system has reached a fixed point. Indeed, small groups (or even an isolated node)
might merge with another one but be unable to reach consensus 
because of too different opinions, and thus split again at the next time step. This might
happen several times in a row, resulting in an evolution of the structure but not
of the opinions.
For this reason, after Eq. \eqref{eq:conv_crit} is satisfied, we let the model run for $10 \times N_{op}$
additional time steps, where $N_{op}$ is the number of different opinions present in the system
at that time.

It is worth noticing that the condition of Eq. \eqref{eq:max_min_condition} rules whether agreement is reached or not by relying only on the most extreme opinions within a group.
This generally makes it hard for large groups to reach consensus.
Although this assumption finds empirical correspondence in some contexts \cite{Hare1952}, in general it might be considered as too restrictive as not taking into account 
the number of agents in $e$ nor the way their opinions are distributed.
To explore this issue, we also consider a different, less strict, rule for group agreement in the Supplementary Material (SM).
This rule, originally proposed in \cite{Hickok2022}, makes agreement possible if the variance of agents' opinions  
within a group is smaller than a given threshold, highlighting the
importance of mediators in group discussions.

\section{Results}

To explore how the combined mechanisms of higher-order interactions and adaptation influence the dynamics of opinions and the structure of interactions, we implement and study the proposed model through numerical simulations.

First, we initialize the model by specifying the structure of interactions, i.e. the hypergraph $\mathcal{H}_0$, and the opinions of the $N$ agents at time $t=0$, i.e. $\{ x_i(0) \}$. For simplicity, here we consider for $\mathcal{H}_0$ an Erd\H{o}s-Rényi (ER) random hypergraph \cite{Schawe2022, agostinelli2025},
and the initial opinions are taken uniformly at random in the interval $[0,1]$.
We remind that an ER random hypergraph is the generalization of ER random graphs: it is defined
by a set of possible hyperedge sizes $\{\mathcal{S}\}$ and by a set of probabilities
$\{p_s\}_{s \in \mathcal{S}}$, which determine for each size $s$ the probability of any group of
that size to be present in the hypergraph (for each $s$, one can equivalently specify instead
$\langle k_s \rangle$, that is, the average number of groups of size $s$ to which a node belongs).
Here, we consider the case of a unique initial group size $s=M$, with an average hyperdegree $\langle k_M \rangle = 10$, and study the cases of $M=2$, $M=4$ and $M=6$.

In the SM we also explore several other initial conditions, both in terms of initial hypergraph structure and of opinion distributions. 
In particular, we consider non-uniform ER hypergraphs and an empirical hypergraph describing face-to-face interactions \cite{sociopatterns, Genois2018} as initial structures of the interactions.
We show that, as the model is adaptive, the dynamical evolution leads to substantial changes in groups' structure, but the initial topology of $\mathcal{H}_0$ does not affect strongly the evolution and steady state of the system. 
In the SM we also consider non-uniform initial distributions of the opinions, showing that this, on the contrary, can have a strong impact on the final outcome of the model.

\subsection{Steady state}
One of the major questions when studying models of opinion dynamics lays on the long-term configuration of the system, and consists in understanding under which conditions or in which parameter range
macroscopic properties such as polarization or consensus emerge from the microscopic 
mechanisms driving individual behavior. Here, the parameter of interest is the confidence $\varepsilon$
\cite{deffuant2000}: we thus analyze how the set of opinions and the structure of interactions
in the steady state depend on $\varepsilon$.

\paragraph{Phase diagram.}
To characterize the final state, we first measure the following quantities, computed in the final state: 
(i) the number of connected components ($N_{cc}$); 
(ii) the relative sizes of the two largest connected components ($S_{cc}/N$); 
(iii) the relative number of groups (hyperedges), with respect to its initial value ($E/E_0$);
(iv) the average and maximum sizes of the groups. 
The two first quantities give information on how opinions are distributed in the final state. 
Indeed, a connected component is by definition composed by individuals sharing the same opinion
(note that the opposite is not true: two different components might a priori correspond to the same opinion). On the other hand, the two latter quantities give structural information on the higher-order topology.

\begin{figure}[t!]
    \centering
    \includegraphics[width=\textwidth]{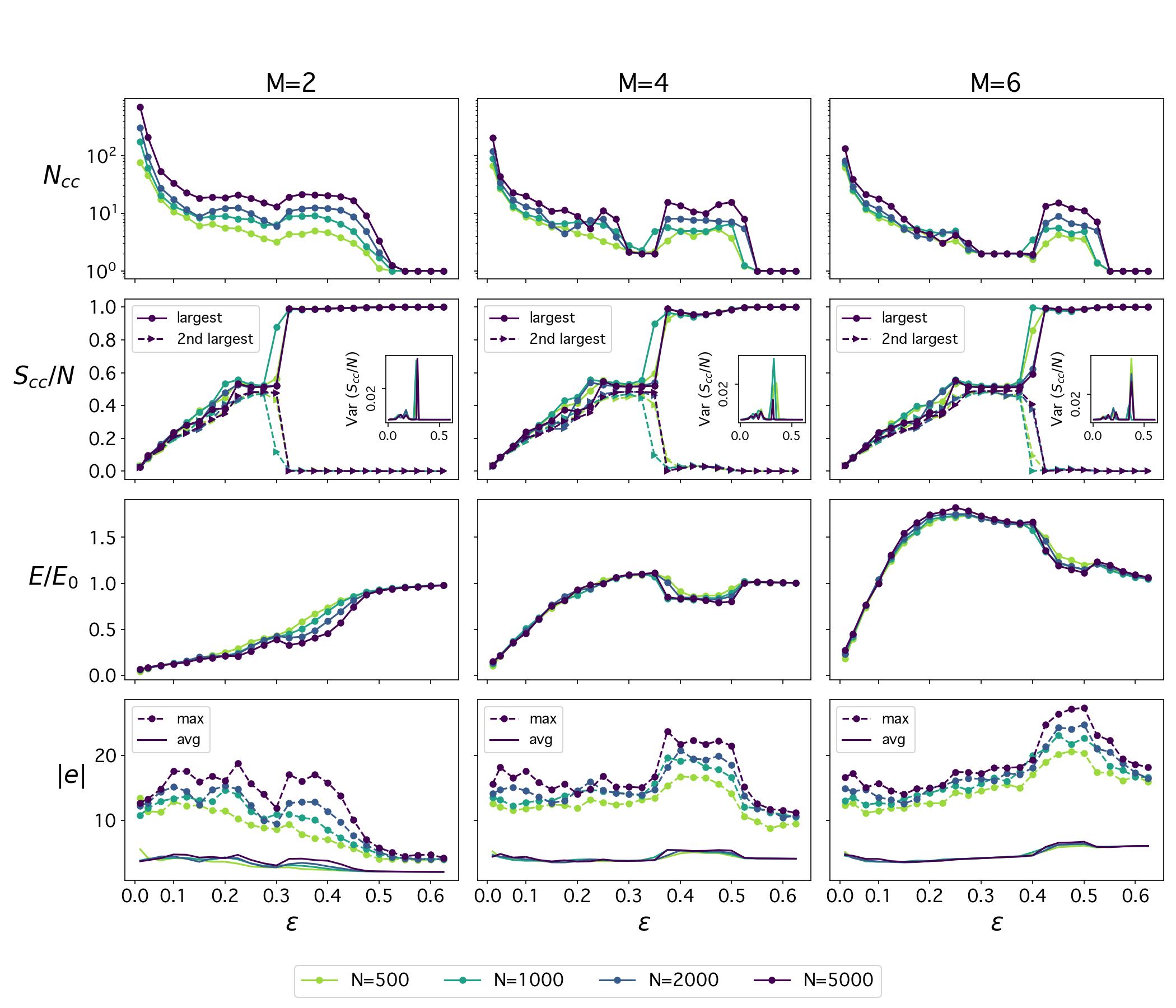}
    \caption{\textbf{Main features of the model in the steady state.}
    Each row shows the behavior of a specific quantity as a function of the confidence parameter $\varepsilon$; columns correspond to different initial group size $M$.
    $\mathcal{H}_0$ is an $M$-uniform ER hypergraph and opinions are initially uniformly distributed in $[0,1]$.
    Top row: number of connected components ($N_{cc}$); 
    second row: relative size of the two largest connected components ($S_{cc}/N$) and variance of the largest one (inset); 
    third row: relative number of hyperedges, with respect to the initial value ($E/E_0$); 
    fourth (bottom) row: maximum (dashed line) and average (solid line) hyperedge size ($|e|$).
    Inside the panels, curves are colored according to the size of the system $N$, from $N=500$ to $N=5000$ (see legend).
    For each value of $\varepsilon$, results are averaged over $20$ independent realizations of the model, 
    starting from different instances of $\mathcal{H}_0$ and $x_i(0)$.
    }
    \label{fig: ER_max_min}
\end{figure}

Figure \ref{fig: ER_max_min} shows how these quantities depend on 
$\varepsilon$, for several system sizes $N$ and initial group sizes $M$.
Let us first focus on the behavior of $S_{cc}/N$ at small $\varepsilon$: although it decreases as $\varepsilon$
decreases, no transition to a fragmented state is observed (a fragmented state is defined as a state in which
no cluster of extensive size is found, hence $S_{cc}/N \to 0$). As observed in \cite{kozma2008} for
the pairwise adaptive Deffuant model, the fragmented phase, which is present in the non-adaptive
case (both for pairwise and for higher-order structures \cite{Schawe2022}),
disappears when rewiring of the topology is enabled: the size of the largest component decays smoothly as the tolerance $\varepsilon$ decreases, but it remains extensive. 
Indeed, rewiring allows individuals with similar opinions but initially far from each other in $\mathcal{H}_0$ to regroup
and create larger clusters instead of remaining isolated, hence preventing a fragmentation regime.

Moreover, we can identify two main regimes in the behavior of $S_{cc}/N$. For increasing $\varepsilon$, the
sizes of the two largest components increase, remaining close until they both reach values close to $N/2$.
This corresponds to a polarized state, in which most of the agents are divided into two connected components.
A plateau is then observed, as the opinions represented in these two components remain incompatible, until
a threshold in $\varepsilon$ is reached: the system
undergoes an abrupt transition to a global consensus, in which a strong majority of agents belongs to the same connected component of size almost $N$, and shares thus 
the same opinion. Interestingly, this transition is present in the pairwise model, both adaptive and
non-adaptive, but was shown to disappear in the Deffuant model with non-adaptive group interactions \cite{Schawe2022} for large enough group sizes, due to higher-order effects.

Overall, the behavior of $S_{cc}$ is very similar to the one observed
in \cite{kozma2008}: adaptivity thus seems to restore the phenomenology hindered by higher-order interactions. 
Contrarily to the non-adaptive higher-order case,
the initial size $M$ of groups does not affect the qualitative behavior 
(absence of fragmentation, and sharp transition between polarized and consensus states), but the transition to consensus is simply shifted to larger values of $\varepsilon$ for larger $M$
(with thus a completely polarized state present on a broader range of $\varepsilon$ values).

While higher-order features do not strongly impact the landscape of opinions of the adaptive Deffuant
model, differences between pairwise and higher-order models emerge at the structural level, as revealed 
by the number and size of hyperedges.
The adaptive dynamics of splitting and merging
of groups can indeed generate groups of various sizes.
As the last row of Fig. \ref{fig: ER_max_min} shows,
sizes much larger than the initial one can be reached,
for any $M$. This effect increases slowly with the system size (for instance, in the SM we show that $|e|_{max}$ grows as $\log N$ for $M=4$): although the initial size of interactions is fixed, larger systems makes it possible for the agents to aggregate into larger groups.
Nevertheless, the average hyperedge size remains approximately stable across different values of $\varepsilon$ and does not vary strongly with $M$ nor with $N$, 
indicating that the typical number of agents interacting in a group is mainly influenced by the rules governing the dynamics, rather than by the parameters of the model.

The various panels of Fig. \ref{fig: ER_max_min} moreover
illustrate an interesting non-monotonous behavior in a range
of values of $\varepsilon$ just above the transition to global
consensus. Although most agents belong indeed to the largest component (with the consensus opinion), the number of connected components $N_{cc}$ presents a range of relatively high values, indicating a prevalence of connected components of small sizes. In the same range, the average and maximum group size increase, while the overall number of groups $E/E_0$ slightly decreases. 
This phenomenon can be understood by considering how the system reaches the steady state, i.e. by focusing on the systems dynamics before the asymptotic regime (that we present in detail in Sect. \ref{sect:temporal}).
At very large $\varepsilon$, consensus is reached easily, so that few split and merge processes are observed, limiting the maximum
group size reached. At values of $\varepsilon$ allowing for a broad consensus but not too large however,
agents with initially extreme opinions might create 
disconnected components more easily; moreover, reaching the steady state can take more time than for large $\varepsilon$, leading to more split and merge processes and allowing the system to create larger groups. Finally, at $\varepsilon$ just below the transition, the system is in the polarized state, and (almost) all initial opinions can converge to one of the two antagonist opinions, thus again limiting the
possibility to create disconnected components.


\begin{figure}[t]
    \centering
    \includegraphics[width=\textwidth]{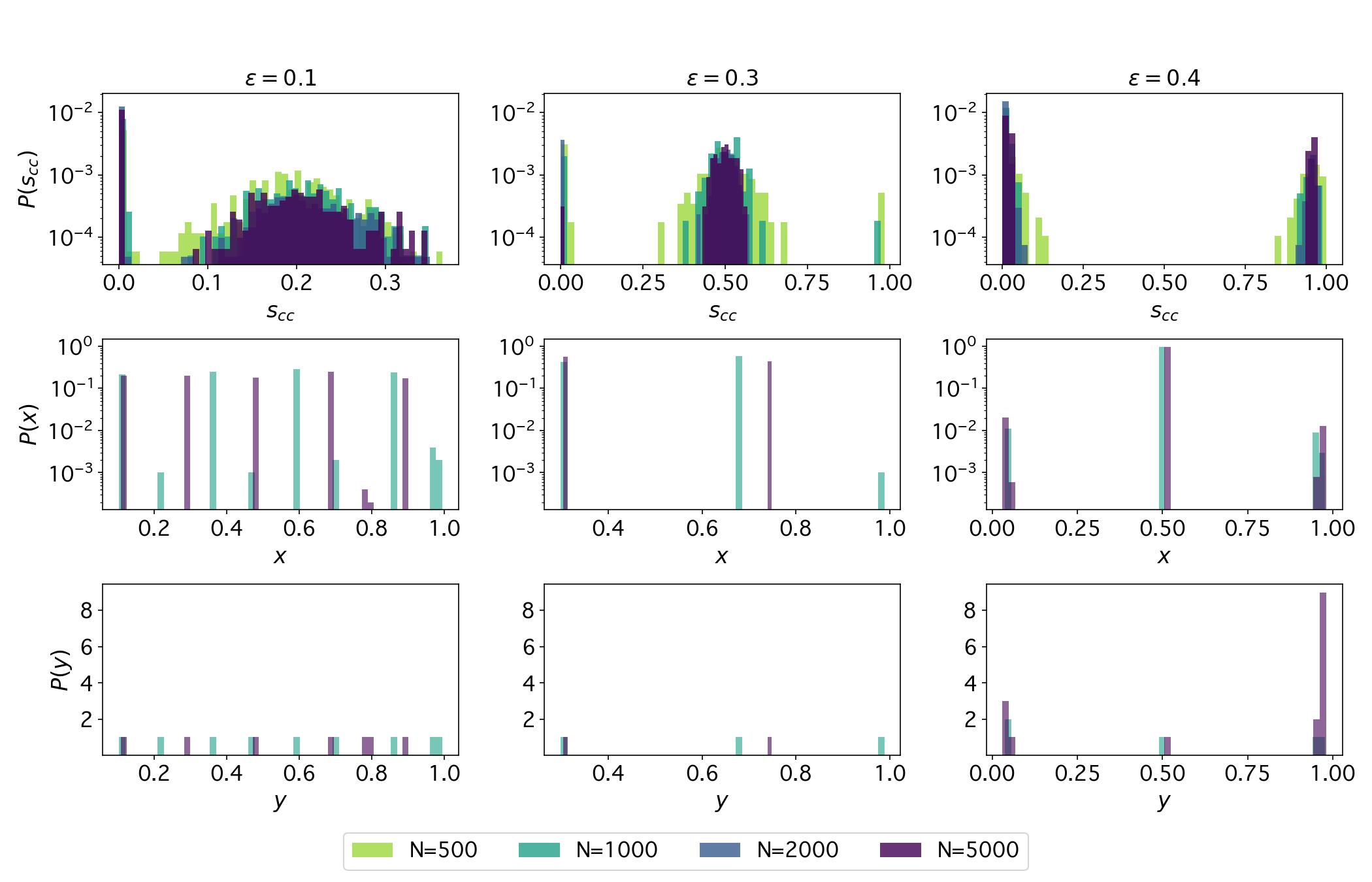}
    \caption{\textbf{Distributions within the steady state, for different values of the confidence parameter $\varepsilon$.}
    First row: distribution of relative sizes of connected components $s_{cc}$.
    Second row: distribution of agents' opinions $x$.
    Third row: number of connected components with opinions $y$.
    The starting structure of interactions is a 4-uniform ER hypergraph (see also central column of Fig. \ref{fig: ER_max_min}) with $\langle k_4 \rangle = 10$, and the initial opinions are uniformly distributed in $[0,1]$.
    Each color corresponds to a specific size of the system (see legend).
    For $P(x)$ and $P(y)$ we show only the values related to one single run with $N=1000$ and one single run with $N=5000$, to improve readability.
    The values of $P(s_{cc})$ are instead averaged over 50 independent realizations.
    }
    \label{fig: ER_max_min_hists}
\end{figure}

\paragraph{Properties of connected components and single agents.}
To characterize the steady state of the model more in detail, we now focus on specific values of $\varepsilon$.
For simplicity, we restrict to the case of $\mathcal{H}_0$ being a 4-uniform ER hypergraph and 
consider the values $\varepsilon=0.1, 0.3, 0.4, 0.6$,
which correspond to four different regimes and behaviors 
as observed in Fig. \ref{fig: ER_max_min}.

For $\varepsilon=0.1$, Fig. \ref{fig: ER_max_min} indicates that the number of connected components is large, the sizes of the largest and second largest 
components are extensive but much smaller than half
of the system size.
To go beyond this simple characterization, Fig. \ref{fig: ER_max_min_hists} explores the distribution of the relative sizes of connected components, $P(s_{cc})$. It shows that the system is organized into few components of extensive sizes and some
very small ones. 
The middle and bottom rows of Fig. \ref{fig: ER_max_min_hists} show respectively the distributions of final opinions of individuals $P(x)$ and of components $P(y)$,
for single runs of the models: 
the large components are of similar 
sizes and represent opinions evenly spaced from each other (of $\approx 2\varepsilon$) as in the original mean-field Deffuant model; some small components of intermediate opinions are present, which have remained isolated during the dynamics (such 
intermediate opinions cannot remain in the mean-field case). Similar results were found in the Deffuant model on adaptive networks without group discussions \cite{kozma2008}, suggesting that higher-order interactions do not influence this aspect of the dynamics.

\begin{figure}[t]
    \centering
    \includegraphics[width=\textwidth]{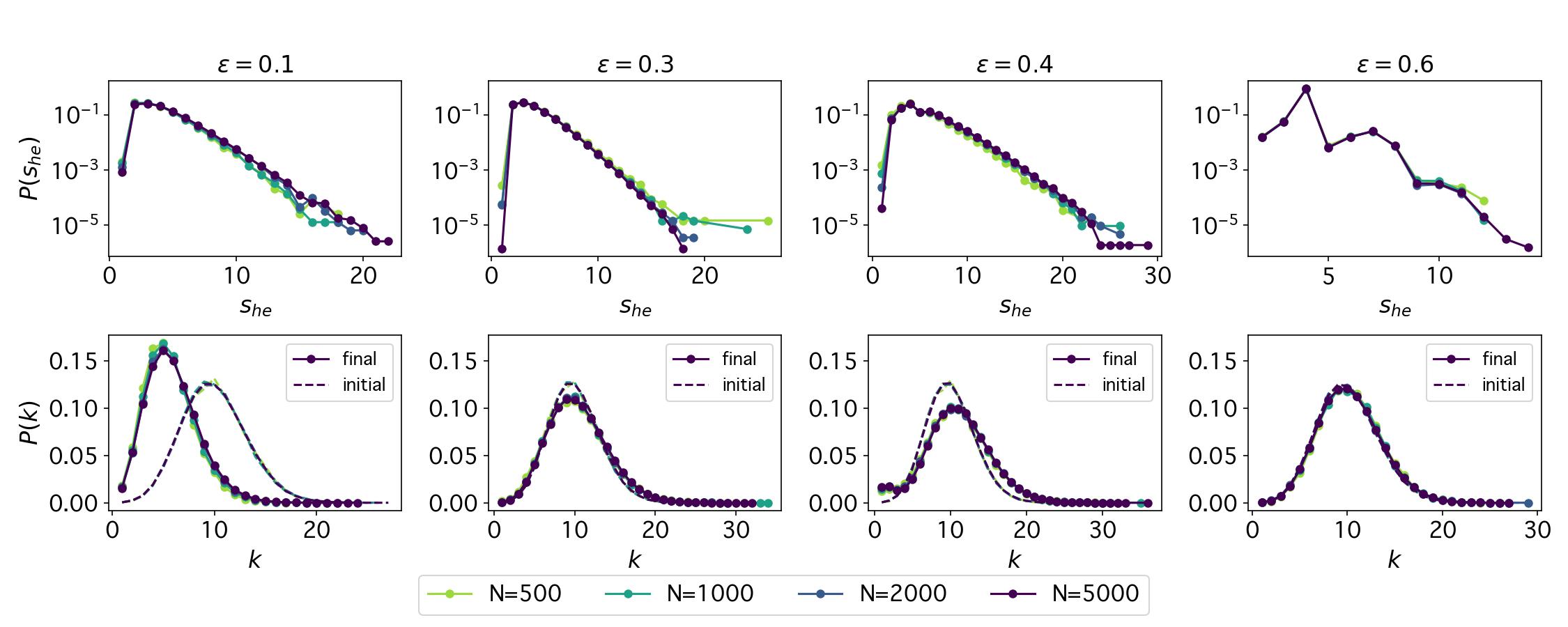}
    \caption{\textbf{Distributions of structural features within the steady state, for four values of $\varepsilon$.}
    First row: distribution of hyperedge (i.e. group) sizes.
    Second row: hyperdegree distribution of the initial (dashed lines) and final (continuous line) hypergraphs.
    The starting structure is a 4-uniform ER hypergraph with average hyperdegree $\langle k_4 \rangle = 10$; initial opinions are uniformly distributed in $[0,1]$.
    Each line color corresponds to a specific size of the system (see legend).
    The results are averaged over 50 independent realizations of the model.
    }
    \label{fig: ER_max_min_distribs}
\end{figure}

For $\varepsilon=0.3$, the system is in the fully
polarized state, with the two largest component
made each by almost half of the population. 
The size distribution of the connected components $P(s_{cc})$ is indeed concentrated around $0.5$, although some very small components can still be present. The distribution of opinions exhibits typically two peaks corresponding to the opinions on which the two components have converged, with occasionally a third peak at an extreme opinion value, corresponding to one or few isolated individuals.

We do not show the results for $\varepsilon=0.6$, as the distribution are trivially Dirac peaks at $s_{cc}=1$,
with all agents carrying the same opinion and being in the same connected component. 
While, even if still in the consensus phase, the case $\varepsilon=0.4$ has a richer phenomenology: in addition to the large cluster of relative size close to $1$, a number of very small clusters are present. Interestingly, the dynamics of opinions has led most agents towards an intermediate value close to $x=0.5$, and the opinions of the small components are instead extreme ones, close either to $0$ or to $1$.
The individuals holding extreme opinions, who have not been able to take advantage of the dynamics to progressively move towards the consensus one, remain thus marginalized and isolated in several small components.


To complete our investigation of the steady state, we focus in Fig. \ref{fig: ER_max_min_distribs} on local structural properties at the various values of $\varepsilon$ considered, studying (i) the distribution of hyperedge sizes $P(s_{he})$ and (ii) of node hyperdegrees $P(k)$ in the final state.
As expected from the dynamical rules of the model and from Fig. \ref{fig: ER_max_min}, $P(s_{he})$ is strongly influenced by group adaptation.
Indeed, in the initial hypergraph $P_0(s_{he})$ is a Kronecker delta $\delta_{s_{he},4}$ by construction,
but in the final state it presents a much broader shape. For all values of $\varepsilon$, small group sizes close to the initial value $M=4$ remain favored, but an exponential tail develops: even if large groups are rare, their sizes can reach values much larger than $M$. This implies that several merging events have managed to create large cohesive groups sharing a common opinion. Note that, as discussed above, the exponential tails extend to larger values for $\varepsilon=0.4$, while the decay is more abrupt for $\varepsilon=0.6$.

\begin{figure}[t]
    \centering
    \includegraphics[width=\textwidth]{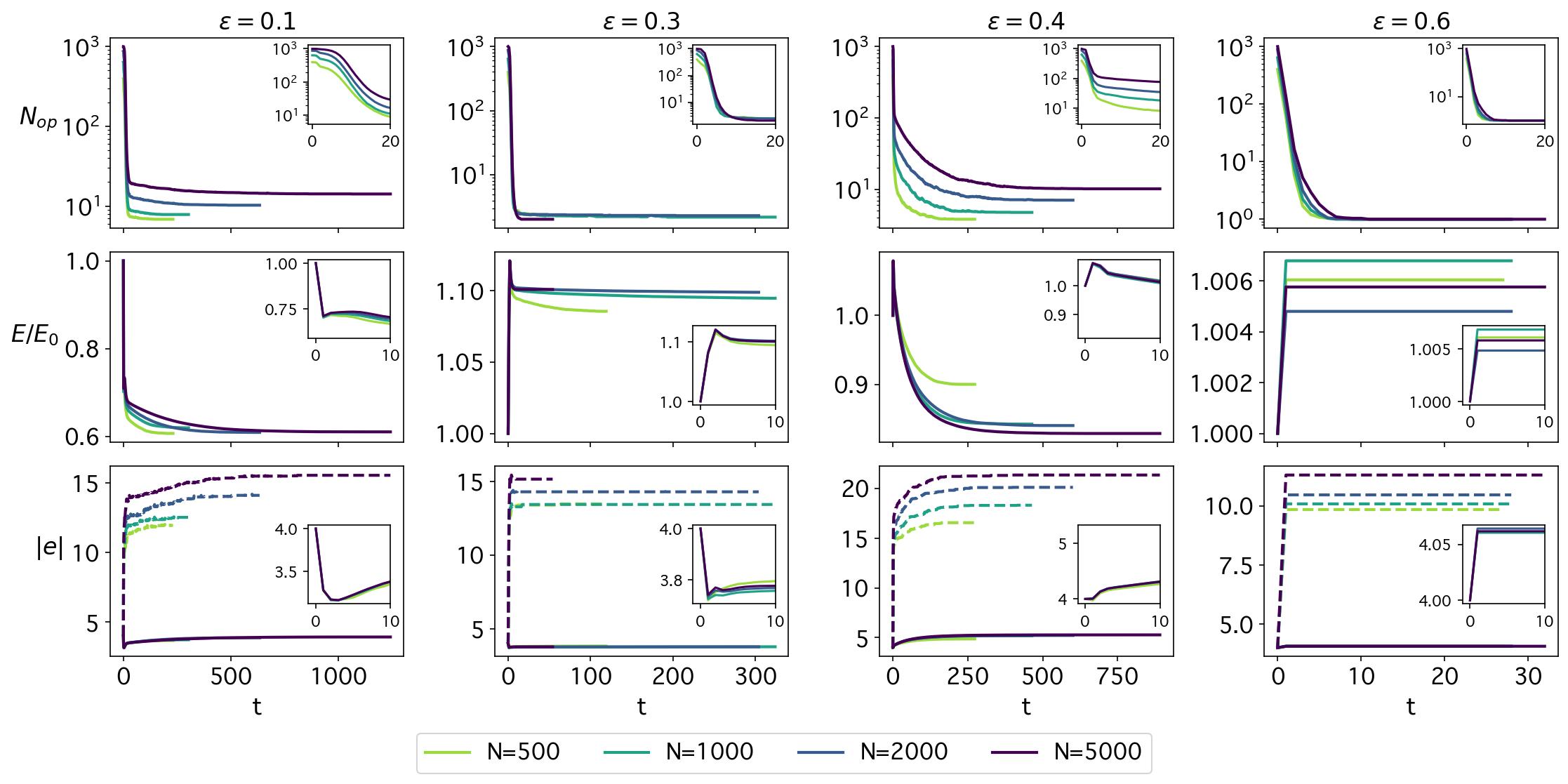}
    \caption{\textbf{Temporal evolution of structural and dynamical features of the system for four values of $\varepsilon$.}
    First row: number of different opinions present in the system ($N_{op}$).
    Second row: relative number of hyperedges ($E/E_0$, w.r.t. the initial value $E_0$). 
    Third row: average (continuous line) and maximum (dashed line) group size ($|e|$). The insets contain the early evolution of the corresponding quantities.
    The starting structure is a 4-uniform ER hypergraph with average hyperdegree $\langle k_4 \rangle = 10$; the initial opinions are uniformly distributed in $[0,1]$.
    The results are averaged over 50 independent runs of the model.
    }
    \label{fig: ER_M4_max_min_temporal_hedges}
\end{figure}

Finally, studying the distribution of the hyperdegree, i.e., of the number of interactions in which an agent participates, gives some further insights.
The initial distribution is by construction a binomial distribution centered in $k=10$, and Fig. \ref{fig: ER_max_min_distribs} (second row) shows how it is impacted by the dynamics. For $\varepsilon=0.6$, the high confidence means that it is easy for an agent to agree with at least part of the other agents in an interaction, or to find another group to interact and agree with. Therefore, the number of interactions of each agent does not change, and the final and initial 
$P(k)$ are superimposed. For $\varepsilon=0.3$ (polarized state) and $\varepsilon=0.4$ (consensus state), the dynamics also impacts only slightly the hyperdegrees: 
$P(k)$ remains close to the initial distribution, and
the initial and final hyperdegrees of agents are
strongly correlated (see SM, where as an example we show that for
$\varepsilon=0.4$ the Pearson correlation between initial and final hyperdegree is $r \simeq 0.85$).
For $\varepsilon=0.1$, the distribution has instead
been largely modified by the dynamics: it is peaked around $k=5$ in the final state. 
This is a direct consequence of the low confidence that makes it more probable for an agent not to agree
with any of the others in a given group, and thus to isolate from that group, decreasing its hyperdegree.

\subsection{Temporal evolution}
\label{sect:temporal}
We now examine how the final state described in the previous section is reached dynamically by the system in the different regimes identified. To this aim, in Fig. 
\ref{fig: ER_M4_max_min_temporal_hedges} we show the temporal evolution of several structural and dynamical features of the system: the number of distinct opinions present in the system $N_{op}$, the number of hyperedges in the structure of interactions $E/E_0$, and the average and maximum hyperedge sizes. 
As in the previous section, we focus on the case of an initial 4-uniform ER hypergraph, and on the values $\varepsilon=0.1, 0.3, 0.4, 0.6$.

Figure \ref{fig: ER_M4_max_min_temporal_hedges} illustrates the presence of different time scales in the temporal evolution of groups and opinions. 
It first highlights that the adaptive dynamics is generally very fast, with the characteristic timescales however depending on $\varepsilon$.
In particular, the first steps see a sharp decrease in the number of different opinions present in the system, corresponding to events of local agreement (first row of the figure). 
In parallel, a first structural adjustment takes place, with the maximal group size increasing fast due to events of splitting followed by merging. 
Moreover, for $\varepsilon=0.1$ the number of hyperedges decreases, while it increases for the other values of $\varepsilon$: 
this is due to the possibility, at small $\varepsilon$, that all initial opinions in the group are far apart, resulting in the total split of the group into isolated nodes (see SM, where we show that for $M=6$ $E$ increases at this value of $\varepsilon$: indeed, for larger $M$ the probability that a group is totally dissolved by this mechanism is lower).
After this fast adjustment of both opinions and structure, a slower evolution takes place (except 
for $\varepsilon=0.6$, since the dynamics is extremely fast as agreement is easy, and no further evolution takes place after the first phase). The evolution is particularly slow for $\varepsilon=0.4$, which is in the range of parameter where a global consensus is reached, but in which a relatively large number of connected components of diverse opinions (extreme ones) remain in the final state, and in which
larger group sizes are also obtained. 
The fact that the dynamics is slower for $\varepsilon=0.4$ than for $\varepsilon=0.3$ is in agreement with the previous discussion concerning this parameter range: at intermediate values of the tolerance, it is easier and faster to reach a polarized state than consensus; in turn, the slower dynamics at $\varepsilon=0.4$ allows to create larger groups. Finally, in the SM we also show the number of events (agreements, splits, and mergers) occurring per time unit, showing again that most of the opinion dynamics occurs in the early time steps, while full convergence takes longer be reached.

\section{Discussion}

In this work, we have presented and studied a bounded confidence model of opinion dynamics, 
which simultaneously takes into account two crucial social mechanisms, only considered separately in previous works
\cite{kozma2008, Schawe2022}. First, agents are not limited to interact in pairs and can discuss in groups of arbitrary size. Second, groups display an adaptive behavior driven by the opinions of their members. 
As in the original Deffuant model \cite{deffuant2000}, a group of interacting agents with similar opinions reaches an agreement, and all the agents of the group converge towards a common opinion \cite{Schawe2022}. 
On the other hand, groups comprising individuals with incompatible opinions are not stable and break into smaller groups, who can then merge with other groups.
The model yields therefore a coupled dynamics of opinions and structure of interactions. Moreover, as interactions are not limited to pairwise ones, contrarily to the adaptive Deffuant model of \cite{kozma2008}, the adaptive dynamics can be richer, and the higher-order topological structure change dynamically.

We have studied by extensive numerical simulations the combined impact of higher-order interactions and adaptivity on the final state of the system and on the system's evolution, both in terms of opinions and structure. 
Our results show the absence of a fragmented state in our model, similarly to the pairwise adaptive case, and in contrast with the non-adaptive one (both with pairwise and higher-order interactions):
even for small values of the confidence parameter, connected components of an extensive number of agents sharing the same opinion can emerge.
Moreover, the transition from polarization to consensus, which has been shown to be smoothed out by higher-order interactions in the non-adaptive Deffuant model \cite{Schawe2022}, is restored by adaptivity.
Thus, similarly to a phenomenon recently observed for spreading processes on adaptive higher-order structures \cite{mancastroppa2026adaptive}, 
the impact of the higher-order nature of interactions is strongly dampened by adaptation. In other terms,
adaptivity, i.e., the ability of agents to restructure their interactions when they cannot agree with others,
appears to be the leading factor driving the global qualitative behavior of the system. 
Of course, group interactions do have an impact, even if limited, on the model's properties. From the structural point of view in particular, 
groups of large size emerge, possibly much larger than the initial groups. 
Even if the initial interactions have a uniform size, the dynamics generates final distributions of sizes with an exponential decay similar to the one observed empirically \cite{Iacopini2024}.
Group structure impacts also slightly the transition between polarization and consensus: indeed including larger groups in the initial structure of interactions increases the corresponding critical confidence parameter value, and a final polarized state is reached for a wider range of confidence values.
Note that, as investigated in the SM, this result is also influenced by the definition of bounded confidence, i.e., by the specific rule for agreement inside groups. Indeed, if the ability of a group
to reach an agreement is determined not by the 
difference between the extreme opinions represented in the group, but by the variance of the opinions,
group agreements are reached more easily, since nodes with intermediate opinions act then as mediators within the group. As a consequence, global consensus is reached for smaller values of confidence (almost independent of the initial group size), and the parameter space that leads the system towards a polarized state is  reduced. 
Finally, results on the dynamics of the model illustrate the presence of different time scales in the temporal evolution of groups and opinions.
The first time steps yield a quick adjustment of 
the number of hyperlinks and of their sizes, as well as a fast reduction of the number of different opinions. In particular, large groups appear already in the early stages of the dynamics. The fast early dynamics is then followed by a slower evolution towards the final state, on timescales depending on the confidence parameter.

Our work brings a new building block in our understanding of complex social phenomena, by 
introducing realistic social mechanisms in a model of opinion dynamics, accounting simultaneously for higher-order interactions and adaptive behaviours and investigating their combined impact. The model's dynamical rules remain however very simplistic and many generalizations and improvements are possible.
For example, the confidence threshold, or the activity level, may be regarded as agent-dependent instead of being constant and uniform for all individuals \cite{Li2023, Li2025}.
Future work could also explore the role of stubborn individuals in shaping consensus, polarization, and fragmentation, as well as the type of transition connecting these phases when higher-order interactions are taken into account \cite{Brooks2024, Centola2018}.
Results in this direction would also contribute to a better understanding of critical mass phenomena in collective behaviors, 
possibly elucidating the role of committed minorities in adaptive systems with group interactions \cite{Iacopini2022,Mancastroppa2023}.
Finally, comparing opinion dynamics models to real-world settings (including experiments using large language models as agents), and defining principled methods to validate them, represent key directions for future investigation \cite{moorsmith2026,Centola2018, centola2015,Ashery2025}.

\section*{Code availability}
The code to reproduce the results presented in this paper is available at \url{https://github.com/cosimoagostinelli/H-or_adaptive_Deffuant}.

\section*{Acknowledgments}
C.A. and A.B. acknowledge support by the “BeyondTheEdge: Higher-Order Networks and Dynamics” project (European Union, REA Grant Agreement No. 101120085).
M.M. and A.B. acknowledge support from the Agence Nationale de la Recherche (ANR) project DATAREDUX (ANR-19-CE46-0008).

\section*{Competing interests}
The authors declare no competing interests.

\bibliographystyle{naturemag}
\bibliography{ref_main}

\end{document}


\maketitle

In this Supplementary Material we consider variations of the model discussed in the main text and present the corresponding results.
In particular, in Section \ref{S1} we assess the impact of a different rule for reaching agreement inside groups.
We also study the impact of different initial conditions regarding either the initial hypergraph structure (Sections \ref{S2}, \ref{S3}) or the initial opinion distribution (Section \ref{S4}).
Section \ref{S5} reports the scaling of the maximum group size in the steady state of the model with the system size; Section \ref{S6} shows the correlation between the initial and final hyperdegree of nodes (i.e. the number of groups in which an agent takes part at the beginning and in the steady state).
Finally, Sections \ref{S7}, \ref{S8} provide more details about the temporal evolution of the model.

\section{Alternative rule for group agreement}
\label{S1}

The agreement rule considered in the main text is commonly used in the literature \cite{kozma2008, Schawe2022}.
However, it represents a strict requirement for groups to converge to a common opinion.
A more flexible alternative was proposed in \cite{Hickok2022}: agents within a group $e$ converge to a common opinion only if
\begin{equation}
    \sqrt{ \frac{1}{|e|-1} \sum_{i \in e} \bigg(x_i - \langle x_j \rangle_{j \in e} \bigg)^2 } < \varepsilon \ ,
    \label{eq:std_conv_rule}
\end{equation}
where $|e|$ is the hyperedge size and $\langle x_j \rangle_{j \in e}$ is the average opinion of the agents in $e$. According to this rule, individuals with intermediate opinions facilitate agreement, playing the role of mediators between agents with extreme opinions.
Contrarily to the case considered in the main text, here larger groups have higher chance to reach agreement, hence we expect global consensus to be favoured by this mechanism.

\begin{figure}[ht!]
    \centering
    \includegraphics[width=\textwidth]{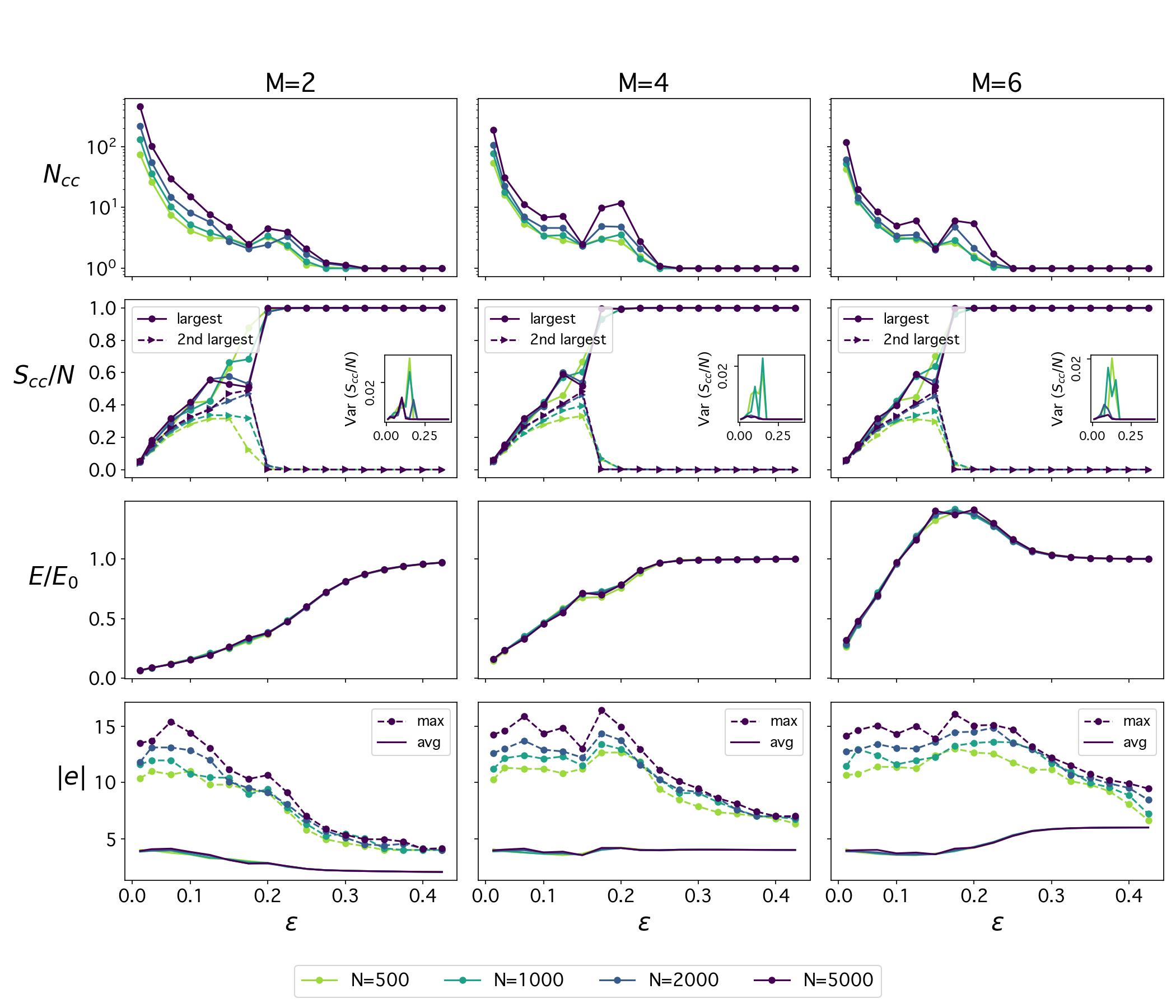}
    \caption{\textbf{Alternative rule for group agreement.}
    Each column corresponds to a different size of hyperedges in the initial ER-hypergraph ($M=2,4,6$).
    Panels in the same row show a specific variable as a function of the confidence parameter $\varepsilon$.
    First (top) row: the number of connected components ($N_{cc}$);
    second row: the relative size of the two largest connected components ($S_{cc}/N$) and the variance of the largest one (in the inset);
    third row: the relative number of hyperedges, with respect to the initial value ($E/E_0$);
    fourth (bottom) row: the maximum (dashed line) and average (solid line) hyperedge size ($|e|$).
    Each color corresponds to a specific size of the system ($N=500,1000,2000,5000$).
    The results are averaged over 20 independent realizations and initial opinions are uniformly distributed on $[0,1]$.}
    \label{fig: ER_std}
\end{figure}

We repeat the analysis presented in the main text using Eq. \eqref{eq:std_conv_rule} as the group agreement rule. Figure \ref{fig: ER_std} shows that global consensus is favoured by this rule, as expected: the critical value of the confidence parameter that separates polarization from consensus is lower than $0.2$, whereas the agreement condition considered in the main text led to a critical threshold of at least $0.3$, depending on $M$.
Another important difference is that here the presence of larger groups in the initial hypergraph does not increase the critical $\varepsilon$: as a consequence, higher values of $M$ do not impact the transition towards polarization.
The rest of the phenomenology appears to be quite similar to the case considered in the main text.

\clearpage

\section{Non-uniform ER hypergraphs}
\label{S2}
\begin{figure}[ht!]
    \centering
    \includegraphics[width=0.8\textwidth]{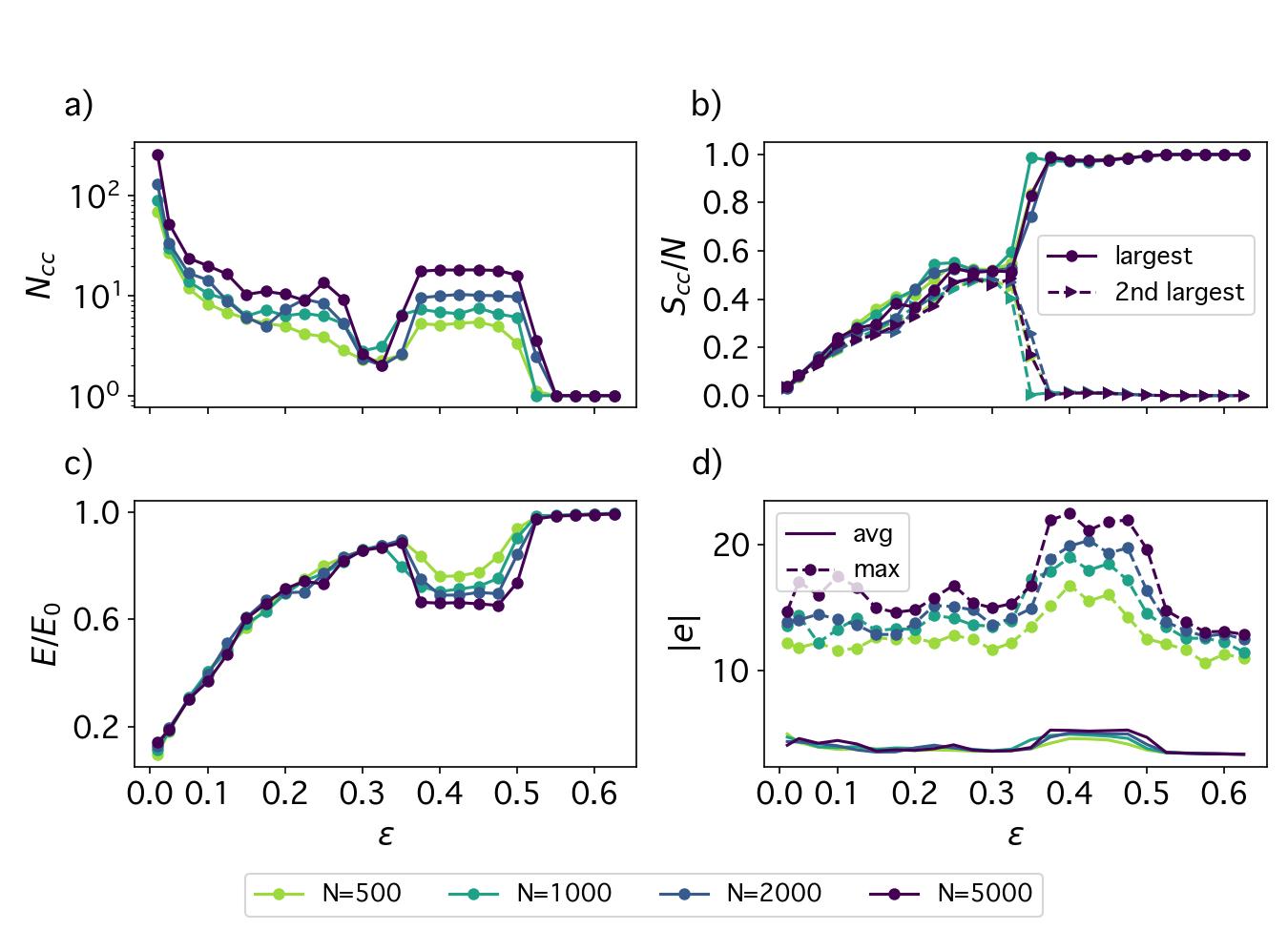}
    \caption{\textbf{Non-uniform ER hypergraphs.} Analogous to Fig. \ref{fig: ER_std}, but we consider as initial hypergraph $\mathcal{H}_0$ an ER-hypergraph with initial hyperedge sizes $s=2,4,6$ and average hyperdegrees $\langle k_s \rangle = 3 \ \forall s \in \{2,4,6\}$.}
    \label{fig: ER_M246_max_min}
\end{figure}

We have shown in the main text that the initial structure of interactions, and in particular the initial group sizes, can impact the long-term configuration of the system.
Here we investigate this point further by considering as initial structure $\mathcal{H}_0$ a non-uniform ER hypergraph, with groups of size 2, 4, 6 and average hyperdegree $\langle k_s \rangle = 3 \ \forall s \in \{2,4,6\}$ (see Fig. \ref{fig: ER_M246_max_min}).
Despite the heterogeneity in the initial group sizes, the steady state of the system resembles the one obtained in the main text, in particular the case of a 4-uniform ER hypergraphs taken as initial structure (see Fig. 2 of the main text, central column).

\clearpage

\section{Empirical hypergraphs}
\label{S3}
So far we have considered ER random hypergraphs as the starting hypergraph $\mathcal{H}_0$ for agents interactions.
Here, instead, we consider an empirical dataset of face-to-face interactions among participants in a conference (SFHH), collected by the SocioPatterns collaboration \cite{sociopatterns, Genois2018}. This dataset was obtained using RFID wearable proximity sensors, which recorded binary interactions resolved in time. A static hypergraph is then obtained with a standard preprocessing procedure from the temporal data: hyperedges are built as fully connected cliques obtained by aggregating interactions that occurred within non-overlapping temporal windows of 15 minutes \cite{agostinelli2025}. The resulting hypergraph consists of $403$ nodes and $6398$ hyperedges, of size up to $10$.

\begin{figure}[ht!]
    \centering
    \includegraphics[width=0.8\textwidth]{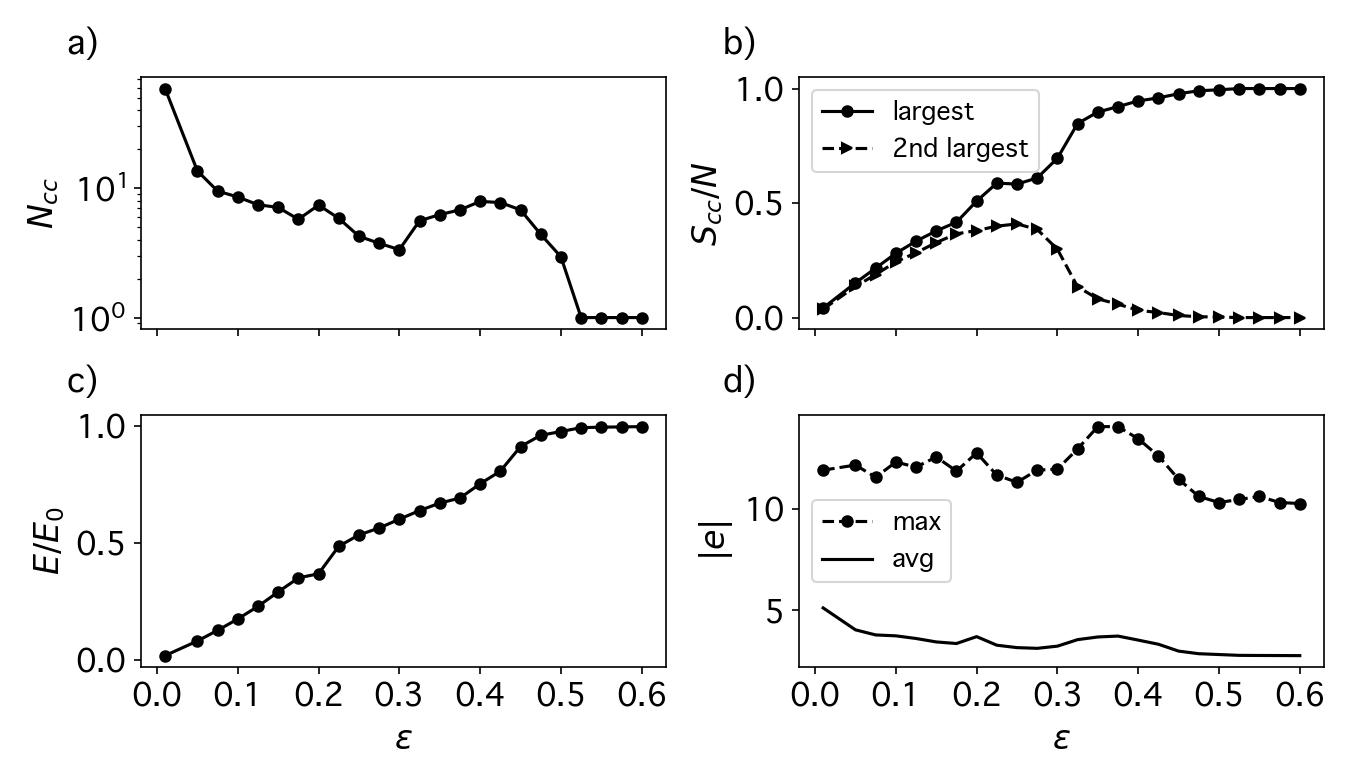}
    \caption{\textbf{Empirical hypergraph - I.} Analogous to Fig. \ref{fig: ER_std}, but we consider as initial hypergraph the empirical hypergraph, SFHH, describing face-to-face interactions in a conference.}
    \label{fig: SP_SFHH_max_min}
\end{figure}

\begin{figure}[t!]
    \centering
    \includegraphics[width=\textwidth]{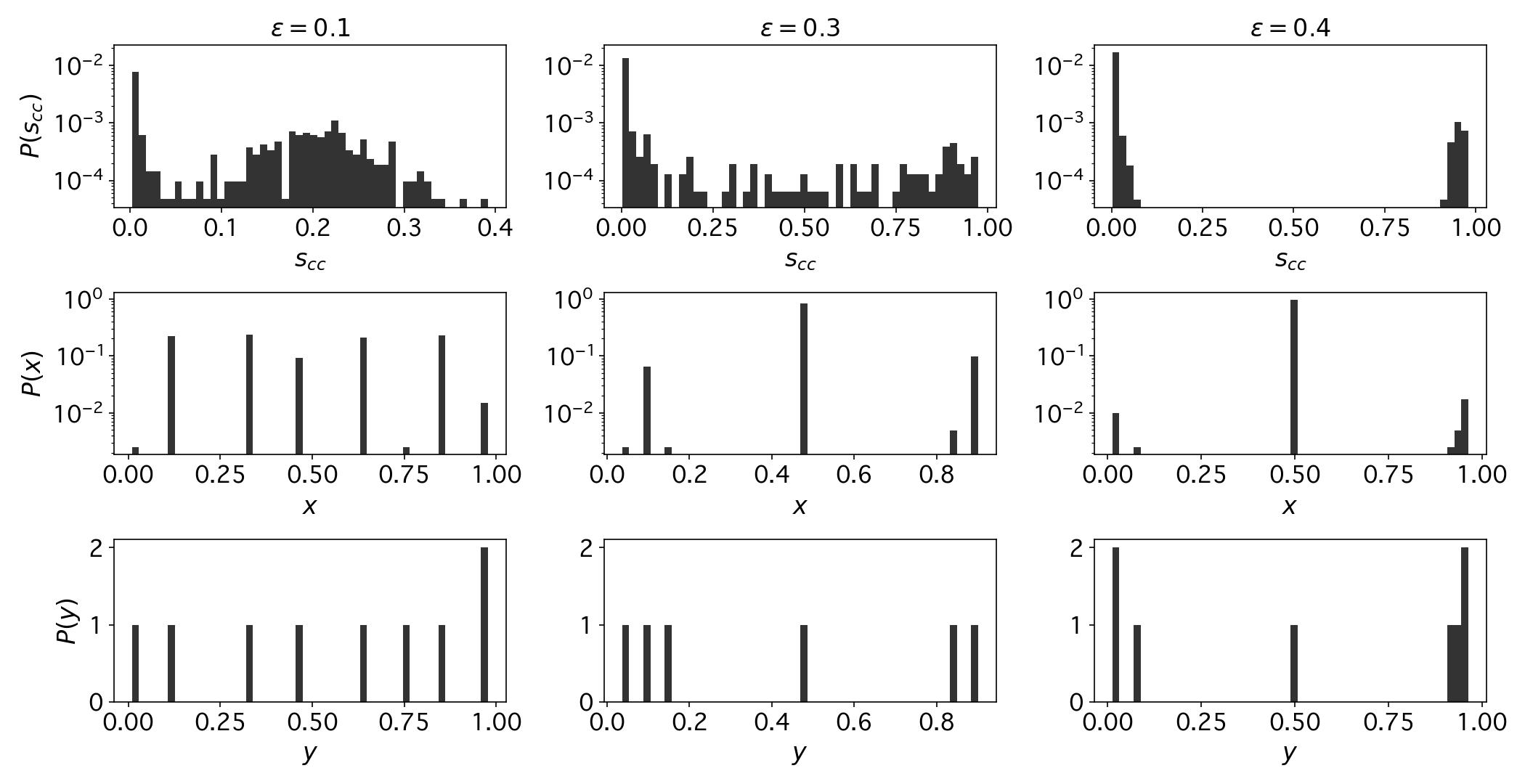}
    \caption{\textbf{Empirical hypergraph - II.} We show three quantities within the steady state of the model, using the SFHH data as starting hypergraph.
    Each column accounts for a value of the confidence parameter $\varepsilon$.
    First row: distribution of relative sizes of connected components $s_{cc}$;
    second row: distribution of agents' opinions $x$.
    third row: number of connected components with opinions $y$.
    The values of $P(s_{cc})$ are averaged over 50 independent realizations, while the others correspond to a single run of the model, in all cases starting with opinions uniformly distributed over $[0,1]$.
    }
    \label{fig: SP_SFHH_max_min_hists}
\end{figure}

\begin{figure}[ht!]
    \centering
    \includegraphics[width=\textwidth]{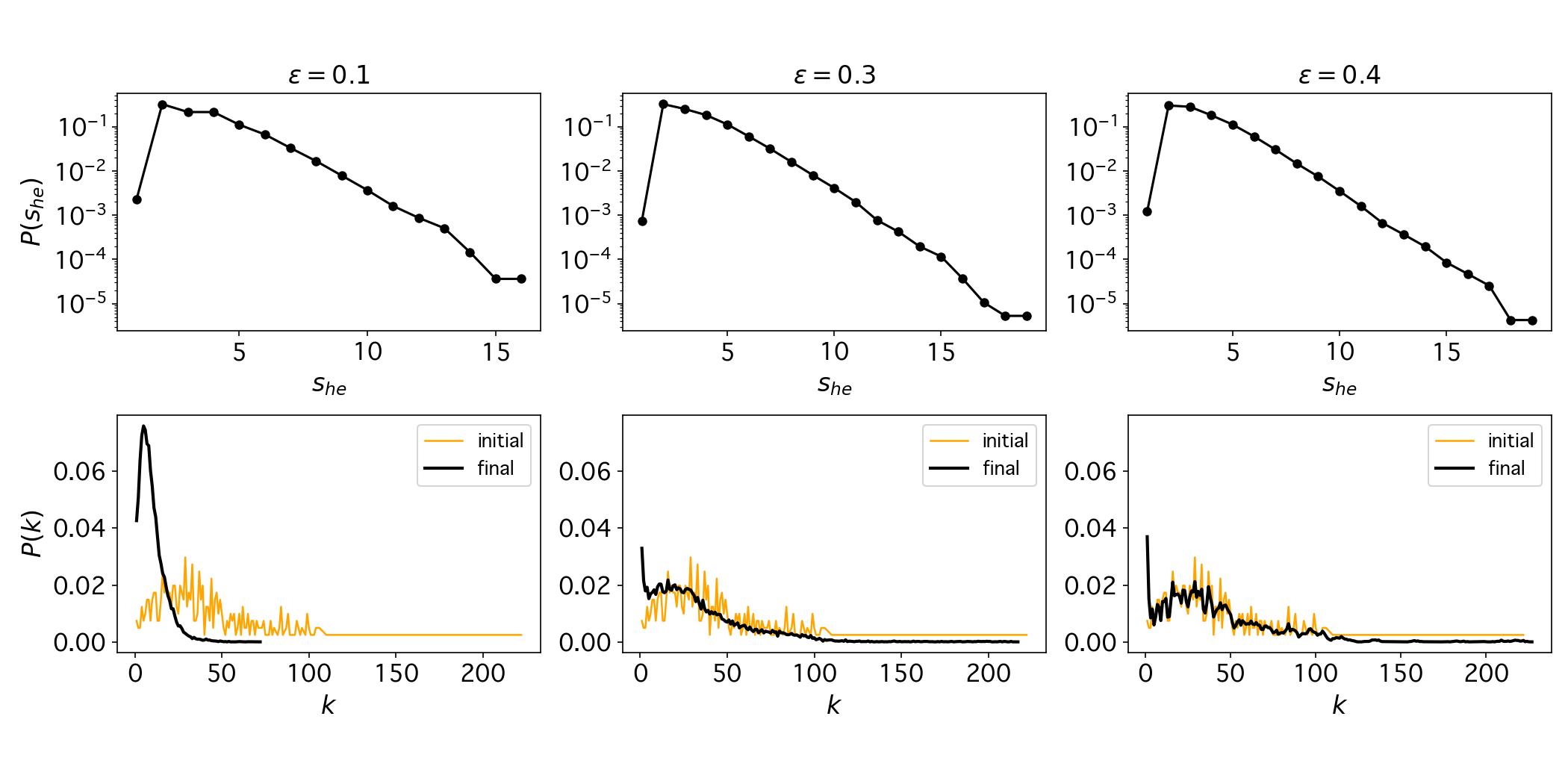}
    \caption{\textbf{Empirical hypergraph - III.} We show distributions related to the system in the final state, starting from the SFHH hypergraph.
    First row: distribution of hyperedge sizes, $P(s_{he})$;
    second row: hyperdegree distribution of the initial (in orange) and final (in black) hypergraph, $P(k)$.
    The results are averaged over 50 independent realizations of the model, starting with opinions initially uniformly distributed over $[0,1]$.
    }
    \label{fig: SP_SFHH_max_min_distribs}
\end{figure}

We employ this hypergraph as initial set of interactions and present the main results of the opinion dynamics model in Fig.s \ref{fig: SP_SFHH_max_min}-\ref{fig: SP_SFHH_max_min_distribs}.
The general picture is very similar to the one presented in the main text when $\mathcal{H}_0$ is an ER hypergraph.
As mentioned in the main text, if $\varepsilon$ is large enough the initial hyperdegree distribution is roughly preserved in the steady state. Contrarily, in a regime of low confidence, individuals cannot sustain a constant amount of interactions and the initial hyperdegree distribution shrinks towards smaller values in the final state.
The slightly different shape of the final hyperdegree distribution with respect to the ER case simply reflects the different topology of the starting hypergraph (Fig. \ref{fig: SP_SFHH_max_min_distribs}, second row).

\clearpage

\section{Bimodal distribution of initial opinions}
\label{S4}
Here, we study how a partial polarization in the initial opinion distribution affects the outcome of the model. To this aim, we consider a bimodal distribution in $[0,1]$, built as the sum of two Gaussian distributions $\mathcal{N}_1(\mu_1, \sigma),\ \mathcal{N}_2(\mu_2, \sigma)$, centered in $\mu_1$ and $\mu_2$ respectively, and with same standard deviation $\sigma$.
We draw the agents' initial opinions from this distribution and study the evolution of the system through the adaptive dynamics we defined.
Note that this initial polarization of opinions does not correspond to a structural polarization, since we still consider ER hypergraphs as starting structure
(agents with similar opinions are not more likely to be connected to each other in the initial state).
There is thus an important difference between a system prepared in this configuration and a system reaching a polarized state after the adaptive dynamics has taken place.

\begin{figure}[hb!]
    \centering
    \includegraphics[width=0.82\textwidth]{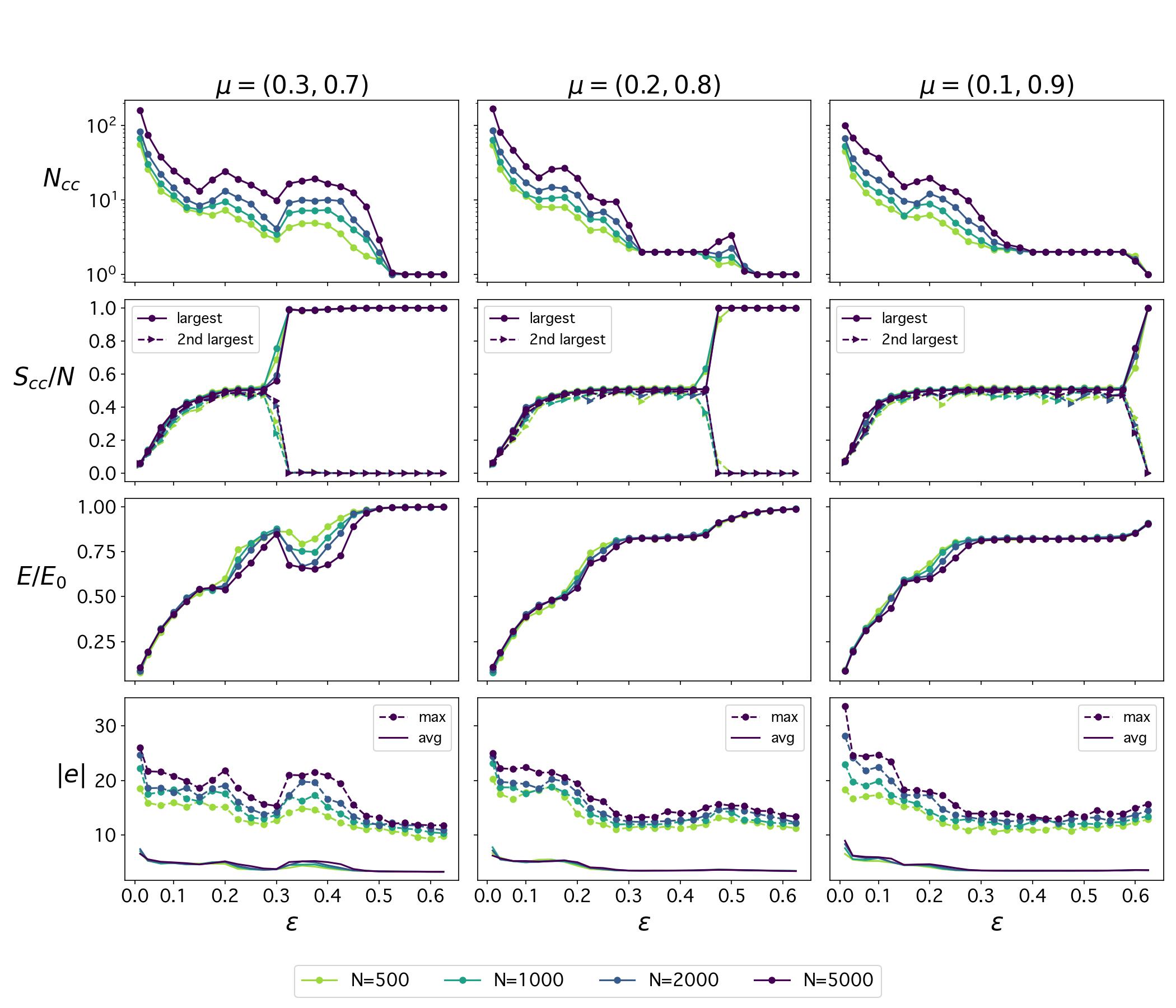}
    \caption{\textbf{Initial bimodal distribution of opinions.}
    Analogous to Fig. \ref{fig: ER_std}, but we consider as initial hypergraph an ER hypergraph with $N$ nodes (see legend), hyperedge sizes $s=2,4,6$ and average hyperdegree $\langle k_s \rangle =3 \ \forall s \in \{2,4,6\}$. Moreover, we consider the initial opinions sampled from a bimodal distribution, obtained as the sum of two Gaussian distributions $\mathcal{N}_1(\mu_1, \sigma),\ \mathcal{N}_2(\mu_2, \sigma)$.
    Each column corresponds to a different pair of means $\mu=(\mu_1,\mu_2)$, while the standard deviation is set to $\sigma = 0.1$.
    The results are averaged over 50 independent runs of the model.
    }
    \label{fig: ER_max_min_bimodal}
\end{figure}

The results show (Fig. \ref{fig: ER_max_min_bimodal}) that, even if the two peaks of the bimodal distribution are not too separated ($\mu_1=0.3,\ \mu_2=0.7$), an almost polarized state is reached 
at smaller values of $\varepsilon$ than when starting from a uniform opinion distribution (see Fig. \ref{fig: ER_M246_max_min}).
Increasing the separation of the two peaks 
 of the bimodal distribution ($\mu_1=0.2,\ \mu_2=0.8$ or $\mu_1=0.1,\ \mu_2=0.9$), leads to a strong extension of the range in which polarization is observed, and a shift of the transition to consensus to larger values of $\varepsilon$: a much higher confidence is needed to reach global consensus when fewer agents with intermediate opinions are present initially, as could be expected.

\clearpage

\section{Maximum group size}
\label{S5}
\begin{figure}[ht!]
    \centering
    \includegraphics[width=0.5\textwidth]{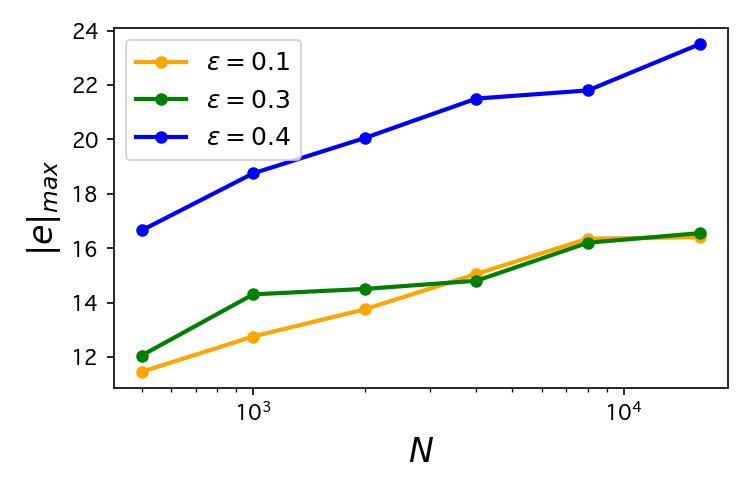}
    \caption{\textbf{Maximum group size as a function of the system size $N$}.
    The initial structures are 4-uniform ER hypergraphs, with $\langle k_4 \rangle = 10$ and the initial opinions are uniformly distributed over $[0,1]$.
    The results are averaged over 20 independent runs of the model.
    }
    \label{fig: e_max_vs_N}
\end{figure}

Figure \ref{fig: e_max_vs_N} shows the dependence of the maximum group size $|e|_{max}$ in the steady state on the number of agents composing the system, for different $\varepsilon$ values.
We take a 4-uniform ER hypergraph for $\mathcal{H}_0$, as a case study.
For all the values of $\varepsilon$ considered $|e|_{max}$ has an approximately logarithmic growth with $N$, consistently with the exponential decrease of the distribution of sizes (Fig. 4 in the main text), indicating that larger systems can give rise to groups composed by a greater number of agents.

\section{Relation between initial and final hyperdegree}
\label{S6}
\begin{figure}[ht!]
    \centering
    \includegraphics[width=0.8\textwidth]{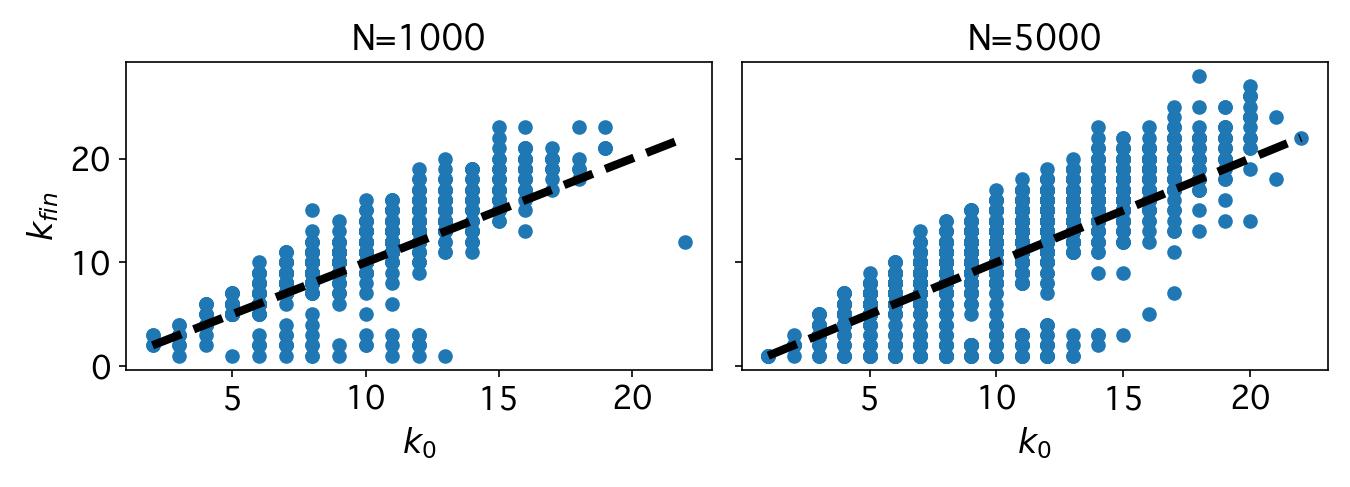}
    \caption{\textbf{Correlation between initial and final node hyperdegree, for $\varepsilon=0.4$}. We show, through a scatter plot, the correlation between the initial and final hyperdegree of nodes, modified by the adaptive dynamics.
    The black dashed line highlights the diagonal.
    The initial structures are 4-uniform ER hypergraphs, with $\langle k_4 \rangle = 10$ and the initial opinions are uniformly distributed over $[0,1]$.
    The results correspond to a single run of the model.
    }
    \label{fig: hdegs_correlation}
\end{figure}

Figure \ref{fig: hdegs_correlation} shows the correlation between the initial and final hyperdegree of every node in the system.
We consider again 4-uniform ER hypergraphs, focusing on two values of $N$ (1000 and 5000) and $\varepsilon=0.4$.
The corresponding Pearson correlation coefficients are: $r=0.86$ (for $N=1000$) and $r=0.85$ (for N=5000).

\clearpage

\section{Temporal evolution for M=6}
\begin{figure}[ht!]
    \centering
    \includegraphics[width=\textwidth]{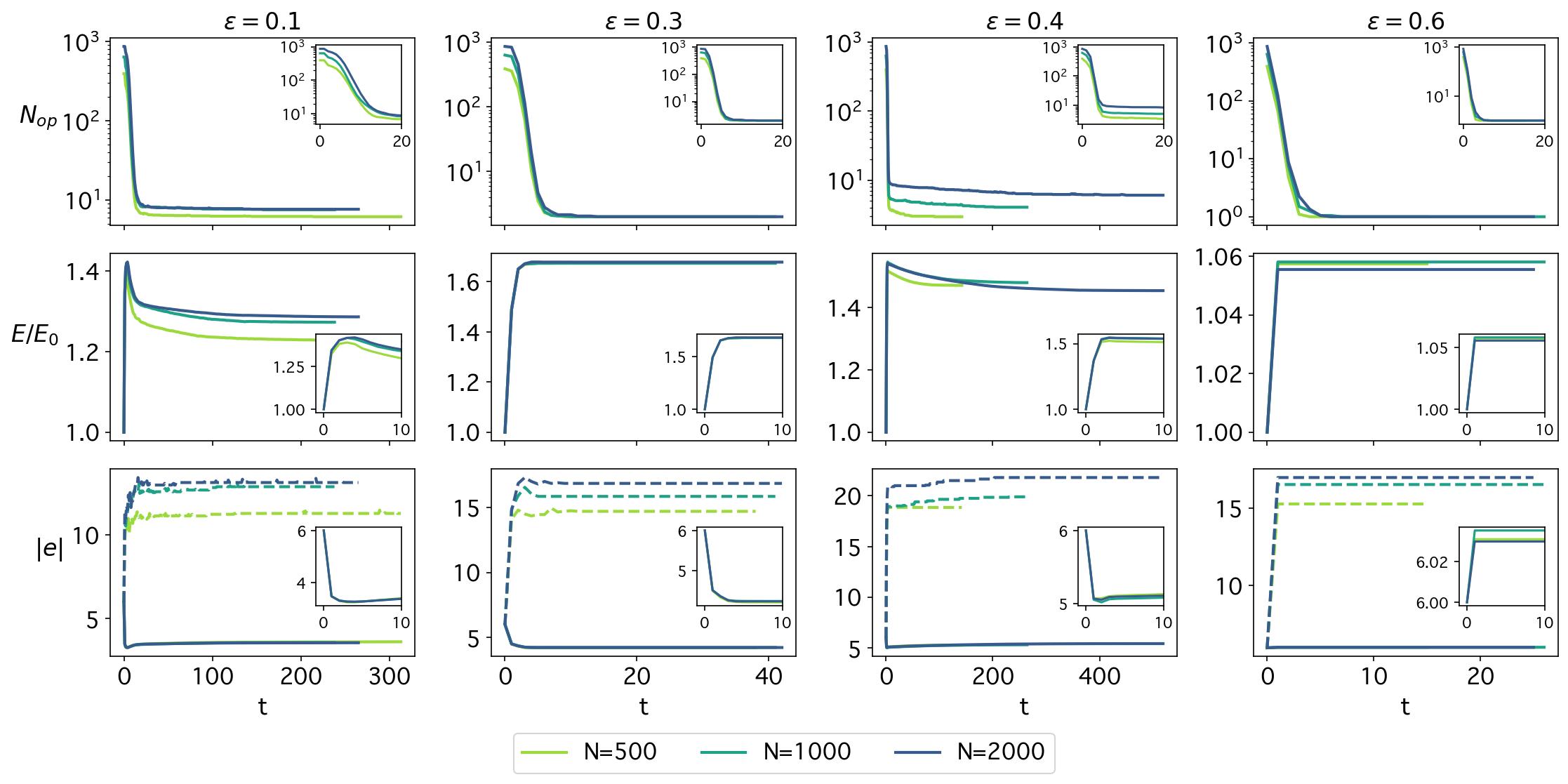}
    \caption{\textbf{Temporal evolution of structural and dynamical features of the system for four values of $\varepsilon$ and M=6.}
    First row: number of different opinions present in the system ($N_{op}$).
    Second row: relative number of hyperedges ($E/E_0$, w.r.t. the initial value $E_0$). 
    Third row: average (continuous line) and maximum (dashed line) group size ($|e|$). The insets contain the early evolution of the corresponding quantities.
    The starting structure is a 6-uniform ER hypergraph with average hyperdegree $\langle k_6 \rangle = 10$; the initial opinions are uniformly distributed in $[0,1]$.
    The results are averaged over 50 independent runs of the model.
    }
    \label{fig: ER_M6_max_min_temporal_hedges}
\end{figure}

Figure \ref{fig: ER_M6_max_min_temporal_hedges} shows the temporal evolution of structural and dynamical features of the system for different values of $\varepsilon$, considering a 6-uniform ER hypergraph
as initial structure $\mathcal{H}_0$ (instead of the 4-uniform ER hypergraph considered in the main text).

\label{S7}

\clearpage

\section{Number of events per time step}
\label{S8}
\begin{figure}[ht!]
    \centering
    \includegraphics[width=\textwidth]{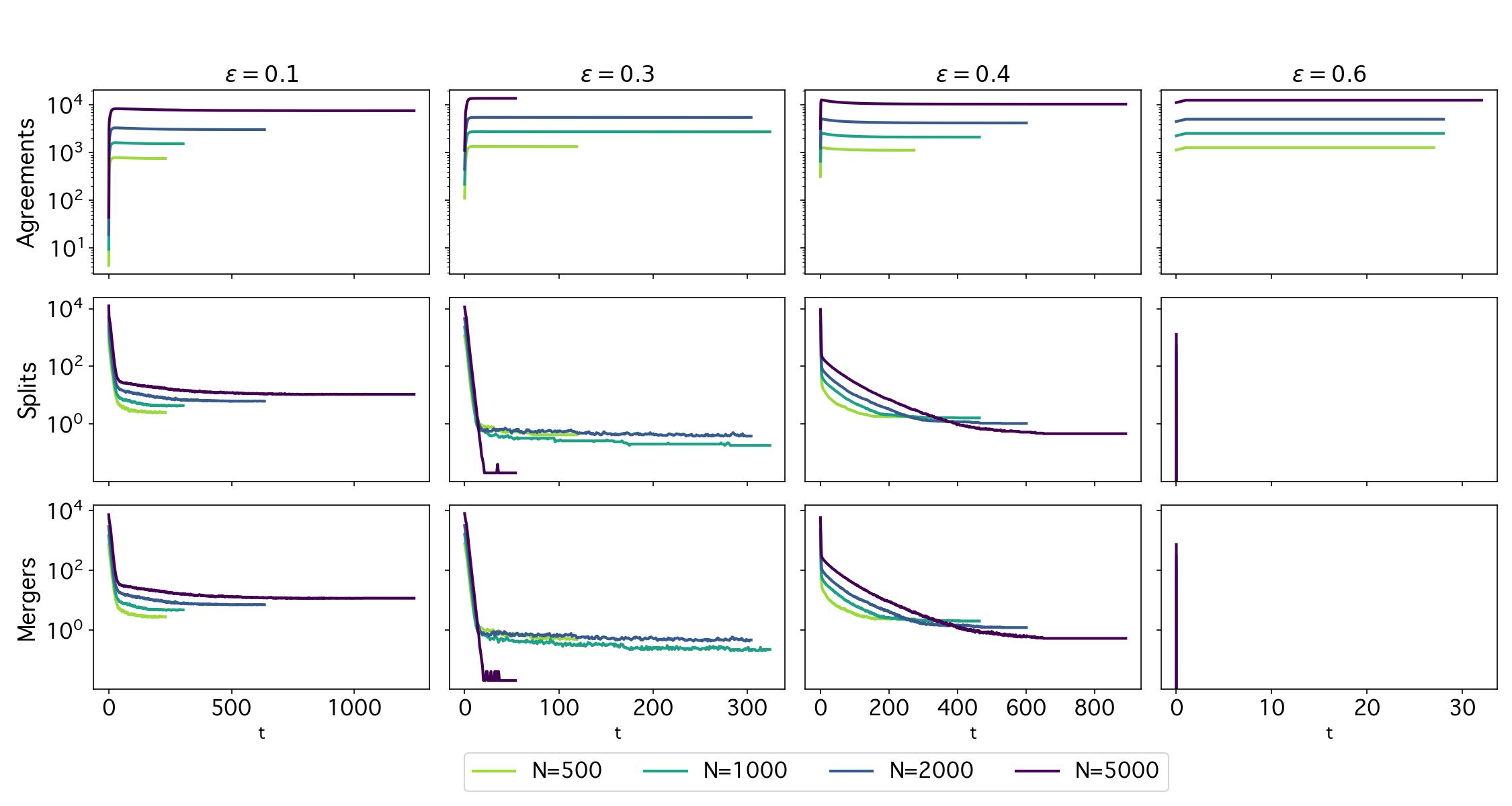}
    \caption{\textbf{Number of events (agreements, splits, and mergers) over time.} We consider four values of $\varepsilon$ (by column) and for each we show the temporal evolution of the number of events (agreements, splits, and mergers) per unit time.
    The starting structure is a 4-uniform ER hypergraph and $\langle k_4 \rangle = 10$; the initial opinions are uniformly distributed in $[0,1]$.
    The results are averaged over 50 independent runs of the model.
    }
    \label{fig: ER_M4_max_min_temporal_n_events}
\end{figure}

To complete the overview about the temporal evolution of the model, we report the number of events (agreements, splits, and mergers) per time unit.
The initial condition on structure and opinions are as in the temporal analysis discussed in the main text.
Figure \ref{fig: ER_M4_max_min_temporal_n_events} confirms the findings previously discussed.
In particular, most of the opinion dynamics occurs in the early time steps, while full convergence takes longer be reached.
Even at the steady state there is a non zero number of split and merger events (the large number of agreements is due to groups who have already converged to the same opinion, and who hence continue to agree at each time step). 
This reflects the presence of isolated nodes (or small groups) that try to find others to agree with, but fail because of their extreme opinion.
Interestingly, in case of a polarized final state ($\varepsilon=0.3$), the largest system converges faster than the others, contrarily to what happens for other $\varepsilon$ values.

\clearpage

\bibliographystyle{naturemag}
\bibliography{ref_supp}